\documentstyle[prl,aps]{revtex}
\large
\begin{document}
\twocolumn
\draft
\title
{Switching Nano-Device Based on Rabi Oscillations}
\author{Thierry Martin$^*$ and Gennady Berman}
\address{CNLS, Theoretical Division, Los Alamos National
Laboratory, Los Alamos NM 87545}
\maketitle


\begin{abstract}
We describe a switching device based on Rabi oscillations. The device
consists of a well region separated from a free region by a potential
barrier. The potential landscape is adjusted
so that one bound state and one quasi level are present. By applying
a microwave field with a driving frequency close to the separation
between the two levels, a particle initially in the ground state
can be activated to the quasi level, and subsequently tunnel to the
free region. The probability for tunneling in the free region
exhibits a plateau structure, as the wave function is emitted by bursts
after each Rabi oscillations.
The leakage current can then be controlled by varying the amplitude of the
external field, the barrier height/width, and the frequency mismatch.
\end{abstract}
\smallskip
\bigskip
\pacs {}
\narrowtext
The recent advances in the fabrication of semiconductor nano-structures
have generated a rebirth of one-dimensional quantum mechanics.
Indeed, while one dimensional (1 D) scattering problems were considered
on a purely academical level in the
sixties, one now has the means to fabricate and design such
1 D structures
using molecular beam epitaxy techniques, and test the prediction
of relatively simple theoretical models.
The motivation to study such artificial structures comes to a large extent
from the hope that these could be ultimately used as elementary
building blocks in the next generation of electronic devices and computers.
Unfortunately,
this research is still at its preliminary stage, and the conceptualization
of even the simplest devices already represents a considerable challenge.
In this letter, we do a step in this direction, and we
consider a device which allows to control
the switching of an electric current: the device exploits the
Rabi oscillations between
a bound state and quasi levels in the presence of a microwave field.
We will restrict the discussion to the single electron
case: electron--electron interactions will be ignored
throughout the paper.

Consider a two state system, with ground state energy $E_G$ and
excited state $E_1$ in the presence of a harmonic perturbation.
If the frequency of the perturbation matches roughly the spacing
between the two levels, the system undergoes oscillations
with a frequency which is much smaller than the excitation frequency
$\omega$.
This Rabi frequency depends on the mismatch $\delta\omega\equiv
(E_1-E_G)/\hbar-\omega$ between the level spacing and the
excitation frequency, and the amplitude, or matrix element
$F_{1G}$ of the perturbation \cite{Landau}:
\begin{equation}
\omega_R=\sqrt{\delta\omega^2+{|F_{1G}|^2\over\hbar^2}}
\label{Rabi frequency}
\end{equation}
If we start with the system initially in the ground state, transitions
to the excited state will occur, but the system will ultimately return
in the ground state after a period $T_R=2\pi/\omega_R$.

We propose to use these oscillations in the following device.
Consider the potential landscape depicted in Fig. \ref{fig1}:
A potential well with depth $V$ is separated from a continuum
region by a thin barrier with height $W$. We adjust the well width $a$
so that there is only one bound state in the well.
The continuum states with $E>0$ extend from the well to the free region.
If the barrier was infinite, we would have a discrete set of states
in the well. For a finite potential barrier, these states become
quasi--levels, which have a finite lifetime in the well.
We choose the barrier parameters in such a way that only one
quasi level exists for the interval $0<E<W$. By applying a dipole field:
\begin{equation}
\phi(x, t)=\epsilon x\cos(\omega t)
\label{dipole}
\end{equation}
with a frequency which is close to the transition between
the ground state and the quasi level, the system will undergo
Rabi oscillations (as long as the wave function has not leaked
totally
out of the well region). By adjusting the amplitude $\epsilon$ of the
external field, the excitation frequency $\omega$ and the barrier
parameters, we can control how much of the wave function
penetrates the free region.

We perform a numerical calculation to probe this device.
In a first step, we need to determine the bound and
continuum states for the potential of Fig. \ref{fig1}.
The ``free'' region is chosen to have a width
$c$ which is much larger than the well and barrier parameters.
The wave functions in all three regions are determined using the
connection formulas for square potential steps, and
the corresponding energies for the bound and ``continuum''
states are found numerically.
To identify the energy around which the quasi level is centered,
we calculate the integrated density of all states with $E>0$,
and select the levels for which this quantity is a maximum.
Alternatively, we calculated all matrix elements of the position operator
between the ground and excited states, and select the level for which
the probability of transition is a maximum
(these two procedures
give the same result).
Once the spacing between the quasi-level energy
$E_Q$ and the
ground state energy $E_G$ is known, we choose the driving
frequency $\omega$ with the desired mismatch
$\delta\omega=(E_Q-E_G)/\hbar-\omega$.

For the time evolution, the dipole potential of Eq. (\ref{dipole})
requires special care. Because of the presence of the position
operator in Eq. (\ref{dipole}),
all terms in the Hamiltonian do not commute with each other,
and the time ordering in the evolution operator \cite{Baym} has to
be considered. However, by performing a gauge transformation
such that the scalar potential in the new gauge is zero and
the driving field is fully included
in the {\it vector} potential, we are able to obtain a Hamiltonian
for which all terms commute (in each constant region of the
static potential $V(x)$): the corresponding vector potential
is constant in space,
and therefore commutes with the momentum operator.
For an elementary time step, the time evolution operator is then
written in the Caley form, and a generalization of the finite
difference scheme of Goldberg, Schey and Schwartz \cite{Goldberg}
for the case of a vector potential is derived.
More details on the method of solution are provided elsewhere
\cite{Martin}.
At $t=0$, the particle is
taken to be in the ground state, and at each time step, we compute
the integrated density in the free region:
\begin{equation}
\rho_i(t)=\int_{a+b}^{a+b+c}d x~\rho(x, t)~,
\label{integrated density}
\end{equation}
as well as the overlap
$|<G|\psi(t)>|^2$ with the ground state.

In Fig. \ref{fig2}a we plot $\rho_i(t)$ for several values of the
driving field amplitude. The barrier width $b=2$ and height $W=3$
are chosen to be large
enough that the ``escape time'' of the wave function is large compared
to other characteristic times of the problem. Moreover,
we have chosen the driving frequency to be close to the resonance condition
($\delta\omega/\omega=0.0001$), which is smaller than the level spacing
for ``continuum'' energies $E>0$, in order to check agreement
with the two level approximation.
For the above parameters, the external frequency is roughly
$\omega=1.5$ in dimensionless units. The infinitesimal
time step $\delta t$ for the numerical evolution has to be
chosen to be small compared to the period of the external field:
here we choose $\delta t=0.0125\times(2\pi/\omega)$.
The integrated density exhibits steps or plateaux,
which allow to identify the Rabi frequency $\omega_R$. Superposed
to the plateaux structure are small amplitude oscillations which period
corresponds to the driving frequency $\omega$. As we start
the simulation, transitions to the quasi-levels and neighboring levels
start occurring, but after a period $T_R=2\pi/\omega_R$, the contribution
of the wave function which remained trapped in the well
has returned for the most part in the ground state. This is
illustrated in Fig. \ref{fig2}b : indeed, aside from a slow decay
associated with the transparency of the barrier, $|<E_G|\psi(t)>|^2$
oscillates with period which matches exactly the plateaux structure.
After the first plateau, $\rho_i(t)$ picks up again as the next oscillation
returns the trapped wave function to the excited states.
The wave function in therefore emitted by ``bursts'' from the well region
every time a Rabi oscillation has been completed.

We note that upon doubling
the coupling strength, the period of the oscillations is halved,
confirming the fact that the Rabi oscillation frequency scales linearly
with the field amplitude if the resonance condition is met.
We compare the measured Rabi frequency with the prediction
for the two level case $\hbar\omega_R=|<G|x|1>| \epsilon$, on resonance.
To estimate the magnitude of the matrix element, we
choose the ground and first excited states of an infinite well,
which yields $<G|x|1>=8a/9\pi^2$. With $a=4$, the predicted Rabi frequency
is therefore $\omega_R^p\sim 0.04$ (dimensionless units).
The measured Rabi frequency yields $\omega_R^m\sim 0.1$, which is
in reasonable agreement with the two level result: the discrepancy
between the two arises from the matrix element estimate.
In addition to these effects, we notice in Fig.
\ref{fig2}a that
after a Rabi oscillation, the integrated density is, as expected,
larger
for higher values of the driving field amplitude.

Next, in Fig. \ref{fig3}a, we compare the evolution of this
system for several values of the barrier widths, keeping the
other parameters fixed. By changing this parameter, we
effectively vary the escape time of the particle
trapped in the well.
For small widths ($b< 1$), the escape
time of the excited wave function is so small that
the Rabi oscillations are hardly noticeable: within a Rabi period,
most of the wave function has tunneled in the free region. The overlap
plot of Fig. \ref{fig3}b confirms these statements, showing a strong
decay over a few Rabi oscillations.
Fig.\ref{fig3}a illustrates how the switching of the current in the
free region can be tuned by varying the width of the barrier.
If we were to further reduce the width of the barrier (say, $b=0.5$), the
Rabi oscillations would cease to be noticeable: the width of the quasi
level becomes large and the (on resonance) matrix element
is strongly reduced in magnitude. Consequently, the activation
mechanism  provided by the dipole potential ceases
to be efficient.

We note that it is also possible to regulate the magnitude of the
leakage current by
varying the off resonance mismatch $\delta\omega$ .
According to Eq. (\ref{Rabi frequency}), changing the mismatch
also changes the Rabi frequency. This is illustrated in Fig.
\ref{fig4}: for the same coupling strength, we compare the
integrated density for mismatchs $\delta\omega/\omega=0.01\%$
(practically on resonance), $\delta\omega/\omega=5\%$,
and $\delta\omega/\omega=10\%$. Away from resonance, the ground state
couples with continuum states neighboring the quasi level
position, which
have smaller matrix elements. Consequently, the leakage
current is smaller than on resonance. Moreover, as
$\delta\omega$ is increased, the first term in the square root
of Eq. (\ref{Rabi frequency}) becomes important, and the period
of the Rabi oscillations is decreased.

We conclude with a brief discussion of the feasibility of this device.
Our main concern is whether phonon relaxation processes
could induce transitions from the quasi level to the ground state
before the excited states can tunnel out of the barrier.
We proceed to convert the parameters of the problem into
dimension full units.
If we choose the height of the barrier to be of the order of $1eV$,
and the effective mass $m^*=0.067m_e$ for GaAs, the width
of the well corresponds to a length $a\simeq 4.8\times 10^{-7}cm$.
The excitation frequency is $\omega\simeq 0.8\times 10^{15}s^{-1}$,
and the constant electric field associated with the dipole potential
is $E\simeq 3\times 10^{5}V/cm$. For GaAs, typical optical phonon
relaxation rates range from $\tau_{ph}\sim 10^{-10}s$ to $10^{-7}$
which is very large compared to all other characteristic length scales
of the problem. We therefore do not expect any significant changes
in the qualitative behavior of the device from these processes.

In summary, we have proposed a switching device based on Rabi
oscillations. The device is composed of a well region with a bound state
and a quasi-level. Upon activation by a microwave field, a particle in the
ground state can be activated to the quasi level, and subsequently
tunnel the free region adjacent to the well. The leakage current can be
controlled in several ways: by varying the intensity of the driving field,
by adjusting the width or height of the tunneling barrier, or by tuning
the resonance mismatch to a desired value. We also note that this device
could in principle also be used ``in reverse'' to measure
the energy of a particle
incident on the barrier and well region, from the free region.
The incident particle tunnels in the well, and in the event that its
energy corresponds to the quasi-level energy, it will get absorbed
in the ground state. If the energy of the incident does not meet
this resonance condition, the wave function should tunnel back
in the free region without noticeable absorption in the ground
state.

\begin{figure}
\caption{Potential landscape: a well region is separated by a
potential barrier from the ``free region''. The well depth and
barrier height are adjusted so that there is only one bound
state in the well, and there is
only one quasi-level below the barrier}\label{fig1}
\end{figure}
\begin{figure}
\caption{a) Integrated density in the ``free'' region as a function
of time, for several values of the coupling constant:
$\epsilon=0.1$ (full line), $\epsilon=0.05$ (dashed line), $\epsilon=0.2$
(dotted line). The time is indicated by the number of infinitesimal
time steps $\delta t=0.0125\times(2\pi/\omega)$ required
for the evolution.
b) Overlap with the ground state as a function of time, with the same
input parameters.}\label{fig2}
\end{figure}
\begin{figure}
\caption{a) Integrated density at fixed coupling strength, but for
different barrier widths: $b=2$ (full line), $b=1.5$ (dashed line),
$b=1.$ (dotted line), $b=0.75$ (dashed dotted line).
b) overlap with the ground state, same parameters.}\label{fig3}
\end{figure}
\begin{figure}
\caption{Integrated density, as a function of time for resonance mismatch
parameters $\delta\omega/\omega=0.01\%$ (solid line),
$\delta\omega/\omega=5\%$ (dotted line), and $\delta\omega/\omega=10\%$
(dashed line)}\label{fig4}
\end{figure}

\begin{thebibliography}{99}
\bibitem[*]{} Present address: Institut Laue--Langevin,
BP 156, 38042 Grenoble, France.
\bibitem{Goldberg} A. Goldberg, H. Schey, and J.L. Schwartz, Am. J. Phys.
{\bf 35}, 177 (1967).
\bibitem{Martin} Th. Martin, preprint (1994).
\bibitem{Baym} G. Baym, {\it Lectures on Quantum Mechanics}, (Addison Wesley,
New York).
\bibitem{Haavig} D. L. Haavig and R. Reifenberger, Phys. Rev. B
{\bf 26}, 6408 (1982).
\bibitem{Landau} L.D. Landau and E.M. Lifshitz, {\it Quantum Mechanics},
(Pergamon, New York 1963).
\end{thebibliography}
\end{document}